\documentclass[useAMS,usenatbib]{mn2e}
\usepackage{graphicx}
\usepackage{lscape}
\usepackage{times}
\usepackage{amsmath}
\usepackage{amsfonts}
\usepackage{amssymb}
\usepackage{multirow}
\usepackage{longtable}
\usepackage{afterpage}
\RequirePackage{color}

\title[Fundamental properties from frequencies]{
Fundamental properties of solar-like oscillating stars from frequencies of minimum  $\Delta \nu$: 
I. Model computations for solar composition
}
\author[M. Y\i ld\i z, Z. \c{C}elik Orhan, \c{C}. Aksoy and S. Ok]{M. Y\i ld\i z$^{}$\thanks{E-mail:
mutlu.yildiz@ege.edu.tr}, Z. \c{C}elik, \c{C}. Aksoy and S. Ok\\
Department of Astronomy and Space Sciences, Science Faculty, Ege University, 35100, Bornova, \.Izmir, Turkey.}

\begin{document}
\date{Accepted 2014 April 3. Received 2014 March 31; in original form 2013 July 27}

\pagerange{\pageref{firstpage}--\pageref{lastpage}} \pubyear{2013}

\def\braket#1{\left<#1\right>}
\newcommand{\numin}{\mbox{\ifmmode{\nu_{\rm min}}\else$\nu_{\rm min}$\fi}}
\newcommand{\numax}{\mbox{$\nu_{\rm max}$}}
\newcommand{\Dnu}{\mbox{$\Delta \nu$}}
\newcommand{\muHz}{\mbox{${\rm \mu}$Hz}}
\newcommand{\kep}{\mbox{\textit{Kepler}}}
\newcommand{\numaxS}{\mbox{$\nu_{\rm max \sun}$}}

\maketitle

\label{firstpage}
\newcommand{\MS}{{\rm M}\ifmmode_{\sun}\else$_{\sun}$~\fi}
\newcommand{\RS}{{\rm R}\ifmmode_{\sun}\else$_{\sun}$~\fi}
\newcommand{\LS}{{\rm L}\ifmmode_{\sun}\else$_{\sun}$~\fi}
\newcommand{\MSbit}{{\rm M}\ifmmode_{\sun}\else$_{\sun}$\fi}
\newcommand{\RSbit}{{\rm R}\ifmmode_{\sun}\else$_{\sun}$\fi}
\newcommand{\LSbit}{{\rm L}\ifmmode_{\sun}\else$_{\sun}$\fi}

\begin{abstract}
Low amplitude is the defining characteristic of solar-like oscillations. 
The space projects $Kepler$ and $CoRoT$  give us a great opportunity to 
successfully detect such oscillations in numerous targets.
Achievements of asteroseismology depend on new discoveries of connections 
between the oscillation frequencies and stellar properties.  In the previous 
studies, the frequency of the maximum amplitude and the large separation 
between frequencies were used for this purpose. In the present study, we 
confirm that the large separation between the frequencies has two minima at 
two different frequency values. These are the signatures of the He {\small II} ionization 
zone, and as such have very strong diagnostic potential. We relate these minima 
to fundamental stellar properties such as mass, radius, luminosity, age and mass 
of convective zone. For mass, the relation is simply { based on the ratio of the
frequency of minimum $\Delta \nu$ to} the frequency of maximum amplitude. 
These frequency comparisons can be very precisely computed, and thus the
mass and radius of a solar-like oscillating star can be determined to high 
{ precision}.  We also develop a new asteroseismic diagram which predicts structural 
and evolutionary properties of stars with such data. We derive expressions for mass, 
radius, effective temperature, luminosity and age in terms of purely asteroseismic 
quantities.  For solar-like oscillating stars, we now will have five very 
important asteroseismic tools ({ two frequencies of minimum $\Delta \nu$, the 
frequency of maximum amplitude,} and the large and small separations between the 
oscillation frequencies) to decipher properties of stellar interior astrophysics.


\end{abstract}


\section{Introduction}
Every object in the universe oscillates in its own way.
For stars, the increasing sensitivity in detecting oscillations in solar-like 
objects by the space missions $Kepler$  and $CoRoT$ is ushering in a new era in 
stellar astrophysics.
Determination of fundamental properties of single stars from oscillation 
frequencies ($\nu$) is among the promises of asteroseismology. 
The relation between the mean density and the large separation between the 
oscillation frequencies (${\Delta \nu}$) is well known (Ulrich 1986).
Kjeldsen \& Bedding (1995) proposed a  
semi-empirical relation between fundamental properties of stars and
the frequency of maximum amplitude  ($\nu_{\rm max}$). 
{ In the present study, we suggest two new frequencies 
($\nu_{\rm min1}$ and $\nu_{\rm min2}$) which, together with $\nu_{\rm max}$ 
and the mean of ${\Delta \nu}$ ($\braket{\Delta \nu}$),
can be used to derive expressions for these fundamental properties.
$\nu_{\rm min1}$ and $\nu_{\rm min2}$ are the frequencies at which 
$\Delta \nu$ is minimum. 
We will show that they have excellent predictive power for stellar 
mass ($M$), radius ($R$) and effective temperature ($T_{\rm eff}$) of single stars, especially if the observations
yield accurate values of $\nu_{\rm max}$.}


 
Kjeldsen \& Bedding (1995) estimate {amplitudes}
of solar-like oscillations and then interrelate $\nu_{\rm max}$ and 
$\braket{\Delta \nu}$ to the stellar mass and radius.
{ Relatively simple expressions for $M$ and $R$ 
 as can be written as functions of 
$\nu_{\rm max}$, $\braket{\Delta \nu}$ and $T_{\rm eff}$ (e.g. see Chaplin et al. 2011):}
\begin{equation}
\frac{M}{{\rm M}_{\sun}}=
\frac{(\numax/\numaxS)^3}{(\braket{\Dnu}/\braket{\Delta \nu_{\sun}})^4}
\left(\frac{T_{\rm eff}}{\rm T_{eff\sun}}\right )^{1.5},
\label{equ:M}
\end{equation}
\begin{equation}
\frac{R}{{\rm R}_{\sun}}=\frac{(\numax/\numaxS)}{(\braket{\Dnu}/\braket{\Delta \nu_{\sun}})^2}\left( \frac{T_{\rm eff}}{\rm T_{eff\sun}}\right )^{0.5}.
\label{equ:R}
\end{equation}
These expressions for $M$ and $R$ are applied in many studies. 
 $Kepler$  and $CoRoT$ data provide us \Dnu ~ and \numax ~for enough stars to 
{ confirm that there is a significant difference 
between the mass found from modelling of these stars and their mass given 
by equation (1) (see, e.g., Mathur  et al. 2012).  
Although a number of papers are dedicated to new scaling 
relations to remove this difference 
(Chaplin et al. 2011; Huber et al. 2011b; Kjeldsen \& 
Bedding 2011; Stello et al. 2011; Corsaro et al. 2013), it is still uncertain if these relations
 are sufficient as written, or are sensitive to other 
stellar parameters not included in them (see, e.g., White et al. 2011).
The scaling relation may depend on, for example, metallicity (see section 5.8). 

For some of the hottest F-type stars, it is reported that 
power envelopes have a flatter maximum (see, e.g., Arentoft et al. 2008; Chaplin \& Miglio  2013). 
For Procyon, for example, photometric and spectroscopic 
methods give different $\numax$ values: 
$\nu_{\rm max,phot}=1014 ~{ \muHz}$, $\nu_{\rm max,RV}=923 ~\muHz$ (Huber et al. 2011a). 
For such extreme stars, it is difficult to use the scaling relations given in equations (1) and (2).
For the stars later than F-type, however, $\nu_{\rm max}$ is much more precisely determined from
the observations than {for} the hot F-type stars. }


Sound speed throughout {a star} changes due to a variety 
of factors. 
{ Abrupt variations can occur due to structural reasons, e.g.,} 
transformation between the energy transportation mechanisms and ionic 
states of certain elements. 
It is thought that such acoustic glitches induce an oscillatory component 
in the spacing of oscillation frequencies
(Houdek \& Gough 2011; Mazumdar et al. 2012). 
{ In particular, changes in physical conditions in the He {\small II} 
ionization zone are very efficient in creating detectable glitches.
Variations of \Dnu ~  around the minima are almost entirely shaped by 
variations of the first adiabatic exponent throughout the zone (see Section 4).}
 

In this paper we suggest two new frequencies and 
show their diagnostic potential by relating them
with the fundamental stellar parameters.
The paper is organized as follows.
{ In Section~2 the basic properties of stellar interior models 
and Ankara-\.Izmir ({\small ANK\.I}) stellar evolution code used in construction 
of these models are presented.
Section 3 is devoted to analysis of oscillation frequencies, the method for 
determination of \numin$_1$ and \numin$_2$, and diagnostic potentials of 
the reference frequencies and their { mode order} differences. 
In Section 4, we consider how the He {\small II} ionization zone influences 
oscillation frequencies and hence their spacing.
Section 5 deals with relating the asteroseismic quantities to the 
fundamental properties of stars.
}
Finally, in Section 6, we draw our conclusions.

\section{Properties of the {\small ANK\.I} code and Models}
\subsection{Properties of the {\small ANK\.I} code }
The models used in the present asteroseismic analysis are constructed by 
using the {\small ANK\.I} 
code 
(Ezer \& Cameron 1965).
Convection is treated with standard  mixing-length  theory  (B\"ohm-Vitense
1958) without overshooting.
{\small ANK\.I} solves { the} Saha equation for hydrogen and helium, 
and computes {the} equation of state by using the Mihalas et al. (1990) 
approach for { survival} probabilities of energy levels
(Y{\i}ld{\i}z \& K{\i}z{\i}lo\u{g}lu 1997). 
The radiative opacity is derived from recent OPAL tables 
(Iglesias \& Rogers 1996), {supplemented} by the low-temperature tables of
Ferguson et al. (2005).
Nuclear reaction rates are  taken from Angulo et al. (1999) and 
Caughlan \& Fowler (1988). 
Although rotating models (Y{\i}ld{\i}z 2003, 2005) and models with 
microscopic diffusion (Y{\i}ld{\i}z 2011; Metcalfe et al. 2012) 
can be constructed by using {\small ANK\.I}, these effects are not included 
in the model computations for this study. 
{ For only the solar model, diffusion is taken into account 
in order to use its known values for the convective parameter 
($\alpha$),  hydrogen  ($X$) and heavy element ($Z$) abundances.}

\subsection{Properties of Models}
Interior models are constructed by using the {\small ANK\.I} code. 
{ The mass range of models is 0.8-1.3 \MS with mass step of 0.05 \MSbit.}
The chemical composition is taken as the solar composition: 
$X=0.7024$ and $Z=0.0172$. 
The heavy element mixture is { assumed to be}
the solar mixture given by Asplund et al. (2009). 
The solar value of the convective parameter $\alpha$ for {\small ANK\.I}~{ is used}: $\alpha=1.98$.  

We have computed adiabatic oscillation frequencies by using {\small ADIPLS} 
oscillation package { (Christensen-Dalsgaard, 2008)} for each mass when the central hydrogen is reduced to 
$X_{\rm c}=0.7$, $0.53$, $0.35$ and $0.17$. 
{We} can compare models with different masses having the 
same relative age ($t_{\rm rel}$).
Define $t_{\rm MS}$ as the main-sequence (MS) lifetime of a star.
Then, for a star having age $t$, relative age becomes 
$t_{\rm rel}=t/t_{\rm MS}$. 
The first value of $X_{\rm c}$ marks essentially the zero-age main sequence 
(ZAMS) age of each 
stellar mass and therefore $t_{\rm rel}$ is very small.  
By definition, $t_{\rm rel}=1$ for terminal-age main sequence (TAMS) models.
The other $X_{\rm c}$ values ($0.53$, $0.35$, $0.17$), however, nearly 
{correspond} to $t_{\rm rel}\approx 0.3$, $0.5$ and $0.75$, respectively.

\textcolor{black}{ In the construction of solar models, diffusion is 
taken into account.}
The maximum sound speed difference between the solar model and the Sun is 
1.7 per cent. 
The base radius of convective zone (CZ) and surface helium abundance 
are 0.732 \RS and 0.25, respectively. 
These values are moderately in agreement with the inferred values from 
solar oscillations: 0.713 \RS and $Y_{\rm s}=0.25$ (Basu \& Antia, 1995; 
Basu \& Antia, 1997). 
Improved solar models (and also models for $\alpha$ Cen A and B) by 
using {\small ANK\.I} are obtained by opacity enhancement (Y{\i}ld{\i}z 2011).

\section{Frequencies of minimum $\Delta \nu$ and their diagnostic potential}
\begin{figure}
\includegraphics[width=97mm,angle=0]{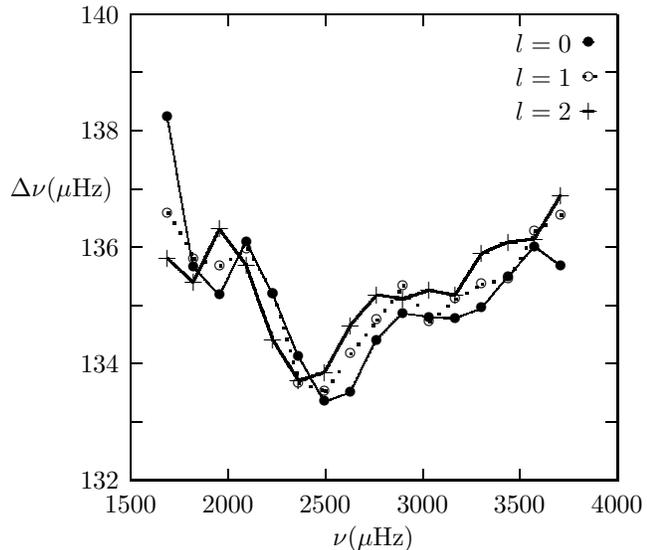}
\caption{ Variation of $\Delta \nu$ as a function of $\nu$ for { the 
BiSON solar data (Chaplin et al. 1999)} 
for $l=0$, $1$ and  $2$.
}
\end{figure}
\begin{figure}
\includegraphics[width=97mm,angle=0]{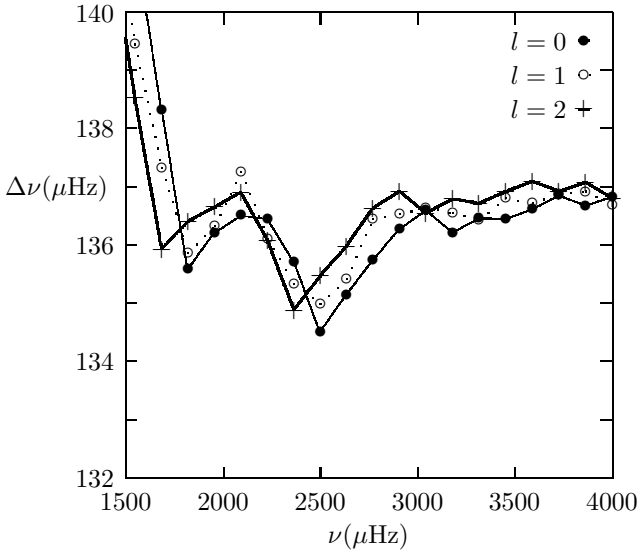}
\caption{ $\Delta \nu$ as a function of $\nu$ for solar model for $l=0$, $1$ 
and  $2$.
}
\end{figure}
\begin{figure}
\includegraphics[width=97mm,angle=0]{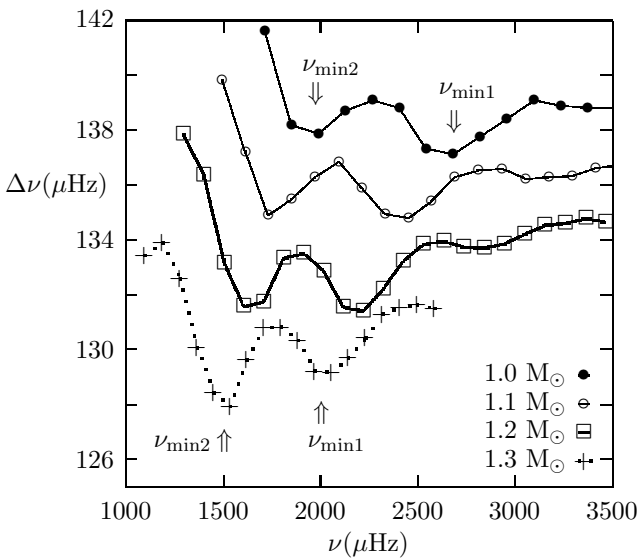}
\caption{ $\Delta \nu$ as a function of $\nu$ for 1.0-1.3 ${\rm M}_{\sun}$ 
models with $X_{\rm c}=0.35$. The degree of the modes is $l=0$ for all of 
the models.
}
\end{figure}
{ 
The asymptotic relation describes the relation between frequency of a mode ($\nu_{nl}$) and its 
{order}  ($n$)  and degree ($l$).
According to this relation, 
the large separation between 
the frequencies ($\Delta \nu= \nu_{nl}-\nu_{n-1,l}$) is to a great extent constant.
This is true for the Sun and other solar-like oscillating stars.
}
{ We compute $\braket{\Delta \nu}$ in the mode order range $n=10-25$.
{
$\Delta \nu$ is plotted with respect to $n$ and a constant function is fitted.
}
For the BiSON solar data (Chaplin et al. 1999), 
$\braket{\Delta \nu}_{\sun}=135.11$ ${\rm \mu}$Hz for $l=0$ and $\braket{\Delta \nu}_{\sun}=135.17$ ${\rm \mu}$Hz for $l=1$.
These results show that $\braket{\Delta \nu}$ is independent of $l$. 
In this study we compute $\braket{\Delta \nu}$ from the modes with $l=0$.
The range of $\Delta \nu$ for degree $l=0$ is 133-138  ${\rm \mu}$Hz}. 
Although this is a very small interval, there { are} 
very significant changes through it.
Variation of $\Delta \nu$ with the frequency is plotted in 
Fig. 1 for $l=0$, 1 and 2. 
The aim of this paper is to make links between such changes and 
stellar parameters. 
The common feature of the three curves is that there are two minima. 
{ We call the minimum having high frequency as the first minimum 
and the other one as the second minimum.
The frequency of the first minimum ($\nu_{\rm min1}$) is around 2600 ${\rm \mu}$Hz, 
and the second ($\nu_{\rm min2}$) is around 1900 ${\rm \mu}$Hz. 
Do these minima also exist in the eigenfrequencies of a solar model?}
In Fig. 2, $\Delta \nu$ of 1.0 \MS model is plotted with respect to $\nu$ 
for three values of $l$ when $X_{\rm c}=0.35$. 
{ We confirm that both of the minima seen in the Sun also exist for the 
oscillation frequencies of 1 \MS model. 
We now consider whether} this kind of variation also appears in 
$\Delta \nu-\nu$ { graphs for other} solar-like 
oscillating stars of different mass.
 

In Fig. 3,{  $\Delta \nu$ is plotted with respect to $\nu$ for 1.0, 1.1, 1.2 and 1.3 \MS interior models with $X_{\rm c}=0.35$.}
{ The eigenfrequencies are for the $l=0$ modes. 
This is the case throughout this paper, if not otherwise stated. 
{ The values of \numin$_1$ and \numin$_2$ for 1.0 and 1.3 
\MS~models are marked in the figure.}
}
As mass increases the minima regularly { shift towards 
lower frequencies.} 
{ While \numin$_1$ is 2600 \muHz ~for the Sun, \numin$_1$ for 
a 1.3 \MS model is about 2000 \muHz.}
\numin$_2$ {for} the Sun is about 1900 \muHz ~and it is 
about 1500 \muHz ~ for the 1.3\MS model. 
The ZAMS model of {a} 1 \MS model has 
\numin$_1$= 3400 \muHz ~ and \numin$_2$= 2500 \muHz.
For interior models of stellar mass up to 1.4 \MSbit, { the minima 
shift again towards lower frequencies
as model evolves { within the MS}.}

\subsection{Determination of \numin$_1$ and \numin$_2$} 
{
For oscillating stars, we have a discrete set of eigenfrequencies. 
In such a case, say, \numin$_1$ does not have to correspond with any of 
the eigenfrequencies.
Then we must determine where the minimum occurs in the $\Delta \nu-\nu$ graph.
Suppose \numin$_1$ is in between the frequencies $\nu_1$ and $\nu_2$.
We use the slopes of the frequency intervals adjacent to $\nu_1$ 
and $\nu_2$ to determine \numin$_1$.
The intersection point of these two lines gives us value of \numin$_1$.
In Fig. 4 two examples for the determination of \numin$_1$ are sketched. 
These are 1.0 \MS models with $X_{\rm c}=0.17$ and $0.35$. }

\begin{figure}
\includegraphics[width=97mm,angle=0]{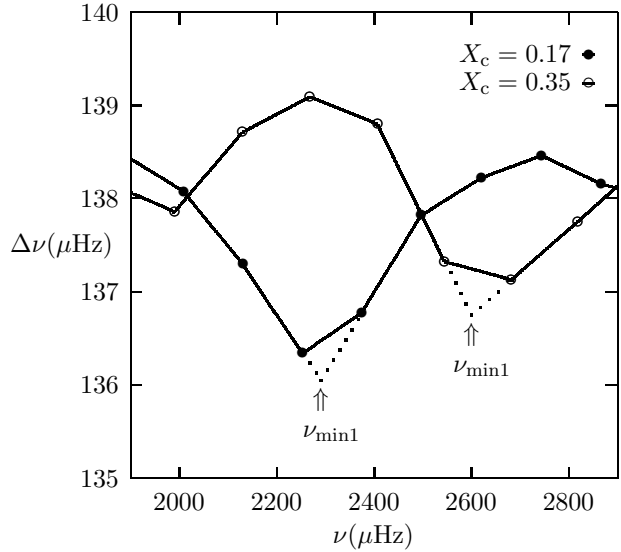}
\caption{ Method for determination of the minima's frequencies. The two examples are 1.0 \MS models with $X_{\rm c}=0.17$
and $0.35$. We first determine frequency interval of the minimum and draw two lines from the 
neighbourhood intervals. The intersection of these two lines gives us \numin$_1$. 
}
\end{figure}
\begin{figure}
\includegraphics[width=97mm,angle=0]{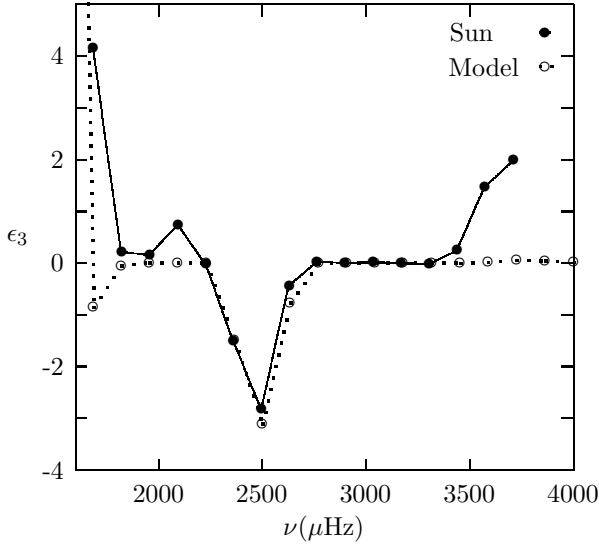}
\caption{ $\epsilon_3$ (equation 3) as a function of $\nu$ for the Sun and solar model 
for $l=0$.
}
\end{figure}

In Figs 1 and 2, it is shown that there are three very similar curves for 
$\Delta \nu$ for different values of oscillation degree 
$l$ ($l=0$, $1$ and $2$).
However, the values of $\nu_{\rm min1}$ for different $l$ are slightly 
different. 
Such a difference may be considered as negligibly small but it may be 
important if one { wants to determine fundamental 
stellar parameters. 
Therefore, we should try to find  a single value for $\nu_{\rm min1}$.
In Fig. 5, the frequency difference parameter $\epsilon_3$ is plotted 
with respect to $\nu$. 
Here, $\epsilon_3$ is defined as
\begin{equation}
\epsilon_3=\prod_{l=0}^2(\Delta \nu_l-{\braket{\Delta \nu}})
\end{equation}
{ where $\Delta \nu_l$ is the large separation for degree $l$.
For each $n$, $(\Delta \nu_0-{\braket{\Delta \nu}})(\Delta \nu_1-{\braket{\Delta \nu}})(\Delta \nu_2-{\braket{\Delta \nu}})$
is computed. $\epsilon_3$ has a much clearer minimum than $\Delta \nu$.
This minimum is the first minimum.
The second minimum for the Sun is missing in Fig. 5 because it is very shallow.
The frequency of the first minimum for the Sun from $\epsilon_3$ is obtained as} 2493.2 \muHz ~for the 
BiSON data. 
However, in some evolved stars, the mixed modes that are 
observed render this method inapplicable. }

\subsection{The diagnostic potential of the { mode order} difference} 
\begin{figure}
\includegraphics[width=101mm,angle=0]{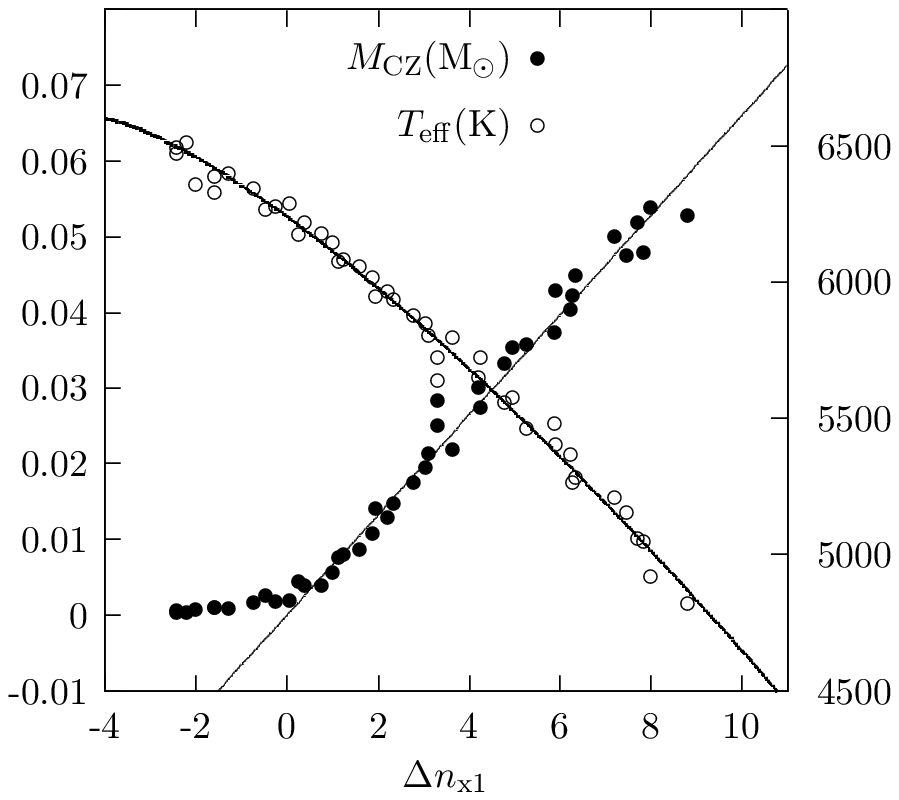}
\caption{  The mass of the CZ and $T_{\rm eff}$ as a function of the { mode order}
difference between \numax ~ and \numin$_1$, 
$\Delta n_{\rm x1}=(\numax - \numin_1)/\Dnu$. 
{
The negative values of $\Delta n_{\rm x1}$ correspond models with $M>1.2 \MS$.
$M_{\rm CZ}$ of these models is very low.
The thin solid line represents the fitted line
for the range of $\Delta n_{\rm x1}=0$$-9$, $M_{\rm CZ}/\MS=0.066\Delta n_{\rm x1}$.
The thick solid line shows the fitted line
for $T_{\rm eff}$ found as $T_{{\rm eff},\Delta n}=(1.142-9.63~10^{-3}(\Delta n_{\rm x1}+4)^{1.35}) {\rm T_{eff \sun}}$.
}
}
\end{figure}
{ 
The relation between the frequencies of two minima is 
{ approximately} given as
\begin{equation}
{\nu_{\rm min1}}\simeq 1.36\nu_{\rm min2}.
\end{equation}
Although { the order of oscillation modes is} not determined 
from observations, the existence of $\nu_{\rm min}$ may solve this problem, 
entirely or in part.
{ Both $\nu_{\rm min1}$ and $\nu_{\rm min2}$ shift { regularly 
as stellar mass and age change.
The difference between $\nu_{\rm min1}$ and $\nu_{\rm min2}$ of the 
models we consider is}
\begin{equation}
\Delta n_{12}=n_1-n_2=(\nu_{\rm min1}-\nu_{\rm min2})/\braket{\Delta \nu}=5-6.
\end{equation}
Its mean value is 5.6. 
The value of $\Delta n_{12}$ is a function of both $M$ and $t$. 
However, the depth of the CZ, 
$d_{\rm BCZ}=(R_\star-R_{\rm BCZ})/R_\star$, is also a function of $M$ and $t$. 
{ Here, $R_\star$ and $R_{\rm BCZ}$ are the stellar radius 
and base radius of the convective zone (BCZ).}
Indeed, there is 
a linear relation between $\Delta n_{12}$ and $1/d_{\rm BCZ}$. 
$\Delta n_{12}$ is about 5 when $d_{\rm BCZ}\approx 0.3$ and 
$\Delta n_{12}$ is about 7 when $d_{\rm BCZ}\approx 0.1$. 

For the mass of the CZ, however, { a stringent relation } is found with 
the { mode order} difference ($\Delta n_{\rm x1}$) between \numax ~ and \numin$_1$.
We define $\Delta n_{\rm x1}$ as 
\begin{equation}
\Delta n_{\rm x1} = (\nu_{\rm max}-\nu_{\rm min1})/\braket{\Delta \nu}.
\end{equation}
Mass of CZ ($M_{\rm CZ}$) is plotted with respect to  $\Delta n_{\rm x1}$ 
in Fig. 6. 
For the models of mass $M<1.2 \MS$, there is a linear relation between 
$M_{\rm CZ}$ and $\Delta n_{\rm x1}$, at least for the MS stars. 
{This relation arises from the fact that both $M_{\rm CZ}$ and $\Delta n_{\rm x1}$ 
are related to $T_{\rm eff}$.
It is } a very strict constraint for interior models of solar-like oscillating 
stars. 
While $M_{\rm CZ}$ of {the} 1.0 \MS~ model with 
X$_{\rm c}=0.35$ is 0.025 \MSbit, the fitting curve gives it as 0.024 \MSbit.
For the MS models of mass $M>1.2 \MS$ ($\Delta n_{\rm x1} < 0$ ), $M_{\rm CZ}$ is negligibly small 
and therefore the method is not applicable.

A similar method can also be obtained for 
$\Delta n_{\rm x2}=(\nu_{\rm max}-\nu_{\rm min2})/\braket{\Delta \nu}$,
the difference between $\nu_{\rm max}$ and $\nu_{\rm min2}$.
The fitting curve for $M_{\rm CZ}$ is 
$M_{\rm CZ}=0.0091\Delta n_{\rm x2}-0.0540$.
For the model given above $M_{\rm CZ}$ is found from $\Delta n_{\rm x2}$ 
as 0.025. 
This result is in { good agreement with $M_{\rm CZ}$ obtained from $\Delta n_{\rm x1}$. 
One can take the mean value of $M_{\rm CZ}$ from $\Delta n_{\rm x1}$ and $\Delta n_{\rm x2}$ }as a constraint to interior 
models of solar-like oscillating stars.
}

{ 
In Fig.6, $T_{\rm eff}$ is also plotted with respect to $\Delta n_{\rm x1}$.
There is an inverse relation between $T_{\rm eff}$ and $\Delta n_{\rm x1}$:
$T_{{\rm eff},\Delta n}=(1.142$$-9.63~10^{-3}(\Delta n_{\rm x1}+4)^{1.35}) {\rm T_{eff \sun}}$.
This relation is very definite and may be used to infer $T_{\rm eff}$ from
asteroseismic quantities alone. The difference between $T_{{\rm eff},\Delta n}$ and model $T_{\rm eff}$ is less than 100 K. 
}

\section{Signature of the H\lowercase{e}{\small II} ionization zone on the 
asymptotic relation}
{
{ The sound speed within a} stellar interior is given 
as $c=\sqrt{\Gamma_1\frac{P}{\rho}}$.
The first adiabatic exponent $\Gamma_1$ is to a great extent constant 
and very close to $5/3$ in the deep solar interior. 
Near the surface, however, an abrupt change in $\Gamma_1$ occurs at 
about $0.98$ \RSbit, as a signature of the He {\small II} ionization zone. 
Such a change significantly influences 
{ the sound speed 
profile near the stellar surface and behaves as an} acoustical glitch for 
the oscillation frequencies.
 
The effect of the acoustical glitch induced by the second helium ionization 
zone on the oscillation frequencies is extensively discussed in the literature 
(see e.g., Perez Hernandez \& Christensen-Dalsgaard 1994, 1998).
In particular, Dziembowski, Pamyatnykh \&  Sienkiewicz (1991), 
Vorontsov, Baturin \& Pamyatnykh (1991), Perez Hernandez 
\& Christensen-Dalsgaard (1994) successfully obtained the helium abundance 
in the  solar envelope from the phase function for solar acoustic oscillations 
(see also Monteiro \& Thompson 2005).}
Houdek \& Gough (2007) consider the second difference as a diagnostic of 
the properties of the near-surface region.
In this section we { consider} how the glitch shapes the variation of \Dnu ~ 
with respect to $\nu$.

The large { frequency separation of a star depends 
on the} sound speed profile in its interior.
It can be written down in terms of  acoustic radius as
\begin{equation}
\Dnu =\left(2 \int_0^R \frac{{\rm d}r}{c} \right)^{-1},
\end{equation}
where acoustic radius $\int \frac{{\rm d}r}{c}$ is the required time for sound 
waves to travel from the centre to the surface. 

As stated above the acoustic glitches induce an oscillatory component 
in the spacing of oscillation frequencies.
Therefore, we are facing a deviation from {the} 
asymptotic relation.
The reason of the oscillatory component is essentially due to coincidence 
of the He {\small II} ionization with the peaks between the radial nodes. 


{Let $\xi_{\rm r}$ be the radial component of the displacement vector.
It gives us the positions of the radial nodes. 
The solution of the second order 
differential equation yields (Christensen-Dalsgaard  2003)
}
\begin{equation}
\xi_{\rm r} = \frac{A}{r(\rho c)^{1/2}} \left| \frac{S_l^2/\omega^2-1}{N^2/\omega^2-1}\right| ^{1/4} \cos
\left(\int_r^{r_2} K(r)^{1/2}{\rm d}r-\frac{\rm \pi}{4}\right),
\end{equation}
where $r_2$ is the outer turning point  and
\begin{equation}
K(r)=\frac{\omega^2}{c^2}\left(\frac{N^2}{\omega^2}-1\right)\left(\frac{S_l^2}{\omega^2}-1\right).
\end{equation}
{ In this equation,} $\omega$ is the eigenfrequency 
obtained from solution of the wave equation. 
$c$, $N$ and  $S_l$ are sound speed, { Brunt-V\"ais\"al\"a} and the characteristic 
acoustic frequencies, respectively. 

\begin{figure}
\includegraphics[width=83mm,angle=-90]{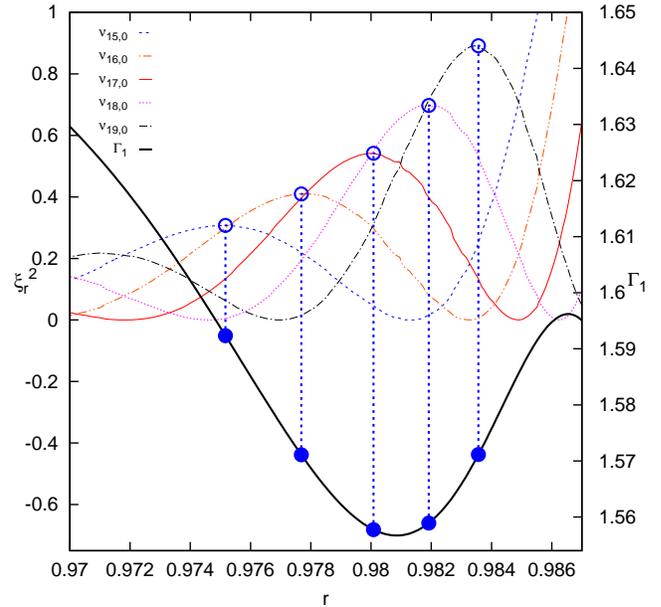}
\caption{ 
{Square of the radial component  (equation 8) 
of {the} displacement vector is plotted against the 
relative radius for the eigenfrequencies of 1.0 \MS model with $X_{\rm c}=0.35$ 
around its \numin$_1$. 
The {amplitude} $A$ in equation (8) is arbitrarily chosen to obtain $\xi_{\rm r}^2$ about unity.
}
Also plotted is the first adiabatic exponent $\Gamma_1$ (thick solid line). 
$\Gamma_1$ has a local minimum about $r=0.98$, due to the He {\small II} ionization zone. 
{ 
The location of the local peak in $\xi_{\rm r}^2$ relative to the dip in $\Gamma_1$
determines the departure of the frequencies from the asymptotic relation, with a
decrease in the frequency that is larger, the closer the peak is to the minimum in
$\Gamma_1$.
}
While the circles show peaks of the oscillations, the filled circles 
represent their projections on $\Gamma_1$. 
}
\end{figure}
The influence of the He {\small II} ionization zone on \Dnu ~ can be understood 
from Fig. 7.
{Square of $\xi_{\rm r}$, given in equation (8), is plotted with respect 
to relative radius around the zone, for eigenfrequencies of the  
1.0 \MS model with $X_{\rm c}=0.35$ near \numin$_1$.} 
Also seen is the first adiabatic exponent $\Gamma_1$. 
The horizontal axis is chosen so that the effect of the zone on $\Gamma_1$ 
is clearly shown. In the  zone,  $\Gamma_1$ has a local minimum where number of 
He {\small II} is the same as number of He {\small III}. 
The largest deviation from the asymptotic relation occurs for the mode 
{ that has one of its peaks closest to the minimum.}
This is the mode with $n=17$. 
{\numin$_1$ is between $\nu_{17,0}$ and $\nu_{18,0}$ (see also Fig. 4).
We note that the minimum of $\Gamma_1$ takes place between the points 
where $\xi_{\rm r}^2$ of modes with $n=17$ and $18$ is maximum.
Therefore, variation of $\Gamma_1$ in the He {\small II} ionization zone shapes 
the variation of \Dnu. 
In order to relate quantitatively the expected local
frequency decrease for specific modes to the minima in \Dnu, further analysis is required.
}}


\section{Stellar parameters from asteroseismic frequencies}
\subsection{The relations between the reference frequencies} 
\begin{figure}
\includegraphics[width=101mm,angle=0]{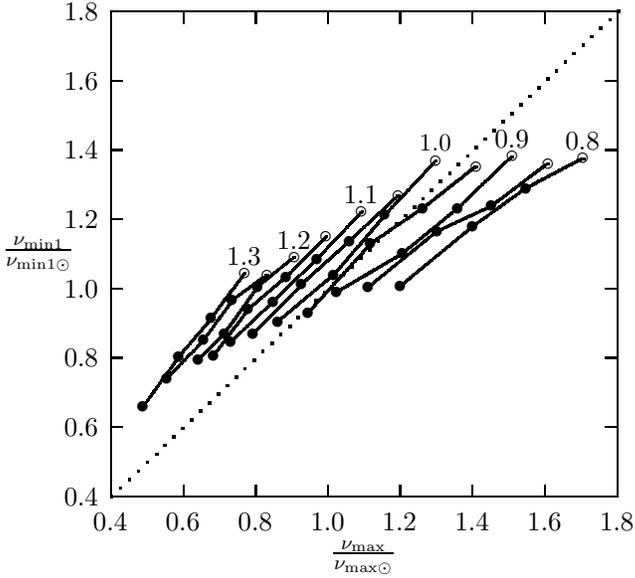}
\caption{  $\nu_{\rm min1}/\nu_{\rm min1\sun}$ with respect to $\nu_{\rm max}/\nu_{\rm max\sun}$.
The circles show the models with $X_{\rm c}=0.7$, while the filled circles 
are for the models with lower values of $X_{\rm c}$. 
The numbers represent the model masses in solar units.  
}
\end{figure}
\begin{figure}
\includegraphics[width=97mm,angle=0]{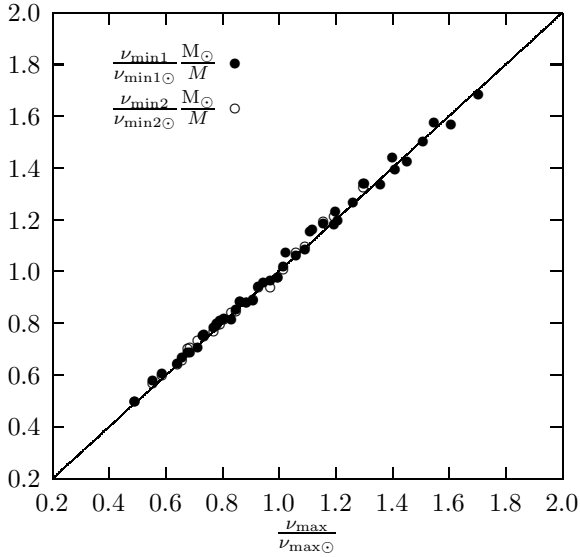}
\caption{ $\nu_{\rm min1}\MS/\nu_{\rm min1\sun}M$ (filled circle) and $\nu_{\rm min2}\MS/\nu_{\rm min2\sun}M$ 
(circle) are plotted with respect to $\nu_{\rm max}/\nu_{\rm max\sun}$.
This shows that $\nu_{\rm min1}\MS/\nu_{\rm min1\sun}M=\nu_{\rm max}/\nu_{\rm max\sun}$.
This equality is a very important tool for computation of stellar mass using asteroseismic methods.
}
\end{figure}
{ 
In our analysis, $\nu_{\rm max}$ of models is computed from equation (1), 
using model values of $\braket{\Delta \nu}$, $T_{\rm eff}$ and $M$.
In Fig. 8, $\nu_{\rm min1}/\nu_{\rm min1\sun}$ is plotted with respect 
to $\nu_{\rm max}/\nu_{\rm max\sun}$. 
The solar values of $\nu_{\rm min1\sun}$ and $\braket{\Dnu_{\sun}}$ are 
found from the BiSON data as 2555.18 and 135.11 \muHz, respectively. 
$\nu_{\rm max\sun}$ is taken as 3050 \muHz.  
{ The dotted line is for 
$\nu_{\rm min1}/\nu_{\rm min1\sun}=\nu_{\rm max}/\nu_{\rm max\sun}$.
They are correlated but there is no one-to-one relation. }
However, we confirm that the difference between 
$\nu_{\rm min1}/\nu_{\rm min1\sun}$ and $\nu_{\rm max}/\nu_{\rm max\sun}$ 
increases as stellar mass is different from 1.0 \MSbit.
The closest models to the dotted line are 1.0 \MS models.
If we plot $\nu_{\rm min1}\MS/\nu_{\rm min1\sun}M$ with respect to 
$\nu_{\rm max}/\nu_{\rm max\sun}$, a linear relation is obtained.
In Fig. 9, $\nu_{\rm min1}\MS/\nu_{\rm min1\sun}M$ (filled circle) is 
plotted with respect to $\nu_{\rm max}/\nu_{\rm max\sun}$.
It is shown that $\nu_{\rm min2}$ obeys the same relation with 
$\nu_{\rm max}$ as $\nu_{\rm min1}$.
The solar value of $\nu_{\rm min2\sun}$ is taken as 1879.52 \muHz, again 
from the BiSON data.
This implies that $\nu_{\rm min1}/M$ and $\nu_{\rm min2}/M$ are equivalent 
to each other. 
Furthermore, they can be used with and without $\nu_{\rm max}$ in new 
scaling relations. 

\subsection{Stellar mass from just ratio of the
frequency of minimum $\Delta \nu$ to
the frequency of maximum amplitude}
A very important result one can deduce from Fig. 9 is that the ratios of 
$\nu_{\rm min1}$ and $\nu_{\rm min2}$ to $\nu_{\rm max}$ are constant. 
The ratio is  independent of evolutionary phase and it simply gives 
stellar mass $M$:
\begin{equation}
\frac{M}{\MS}=\frac{\nu_{\rm min1}}{\nu_{\rm min1\sun}}\frac{\nu_{\rm max\sun}}{\nu_{\rm max}}
             =\frac{\nu_{\rm min2}}{\nu_{\rm min2\sun}}\frac{\nu_{\rm max\sun}}{\nu_{\rm max}}.
\end{equation}
This implies that we can obtain stellar mass in two new ways: one is 
with $\nu_{\rm min1}$ and the other is with $\nu_{\rm min2}$. 
{ 
The masses computed from equation (10) in terms of $\nu_{\rm min1}$ ($M_1$) and $\nu_{\rm min2}$ ($M_2$)} 
are listed in Table 1. 
Equation (10) is a very simple and a new contribution to asteroseismology 
of solar-like oscillating stars.
It is independent of stellar age, at least for the MS evolution. 
That is to say the fractional changes of $\nu_{\rm min}$ 
and $\nu_{\rm max}$ in time are the same. 
However, the effect of chemical composition ($X$ and $Z$) and the 
convective parameter on equation (10) should be tested. 
Such a test is {the} subject of another study.

As seen in equation (10), ${\nu_{\rm min1}}/{\nu_{\rm min1\sun}}$ and 
${\nu_{\rm min2}}/{\nu_{\rm min2\sun}}$ are equivalent to each other. 
{ Therefore, we hereafter prefer to use ${\nu_{\rm min1}}$ only 
but ${\nu_{\rm min2}}$ can also be used provided that 
it is divided by the solar value. }
 
\begin{table}
\caption{Masses and radii computed by using asteroseismic methods. 
They are in the solar units.
$M_{\rm mod}$ and $X_{\rm c}$ in the first and second columns are the 
model mass and central hydrogen abundance, respectively. 
$M_1$ and $M_2$, given in the third and fourth columns, are masses 
computed by using $\nu_{\rm min1}$ and $\nu_{\rm min2}$ (equation 10), 
respectively. 
{ $M_2$ of some low-mass models is absent  because the second minimum is not seen  in the 
eigenfrequencies of these models.}
$M_{\rm e12}$ is computed from equation (12), and $M_{\rm sis}$ is 
obtained from equation (15) or (16). 
$\overline{M}$ given in the seventh column is the mean of $M_1$ and $M_2$. 
We give the percentage difference between $M_{\rm mod}$  and $\overline{M}$ 
in the eighth column. 
In the last two columns, model radius and radius derived 
($R_{\rm e17}$) from $\nu_{\rm min1}$ and $\braket{\Dnu}$ 
(equation 17) are listed. 
}
\centering
\small\addtolength{\tabcolsep}{-3pt}
\begin{tabular}{cccccccccc}
\hline
 $M_{\rm mod}$& $X_{\rm c}$  & $M_1$   & $M_2$ & $M_{\rm e12}$&$M_{\rm sis}$&  $\overline{M}$ & $\delta M(\%)$ & $R_{\rm mod}$& $R_{\rm e17}$ \\
\hline
   0.80 &  0.70 &    0.79 &    --- &    0.84 &    0.79 &    0.79 & $~$ 1.4 &    0.72 &    0.72 \\
   0.80 &  0.53 &    0.81 &    --- &    0.85 &    0.81 &    0.81 & -1.7 &    0.75 &    0.76 \\
   0.80 &  0.35 &    0.81 &    --- &    0.84 &    0.80 &    0.81 & -1.7 &    0.78 &    0.79 \\
   0.80 &  0.17 &    0.82 &    --- &    0.84 &    0.80 &    0.82 & -2.5 &    0.84 &    0.86 \\
\hline
   0.85 &  0.70 &    0.83 &    --- &    0.87 &    0.83 &    0.83 & $~$ 2.7 &    0.75 &    0.75 \\
   0.85 &  0.53 &    0.83 &    --- &    0.87 &    0.83 &    0.83 & $~$ 2.1 &    0.79 &    0.79 \\
   0.85 &  0.35 &    0.87 &    --- &    0.89 &    0.86 &    0.87 & -2.8 &    0.83 &    0.84 \\
   0.85 &  0.17 &    0.88 &    0.87 &    0.89 &    0.86 &    0.88 & -3.3 &    0.89 &    0.90 \\
\hline
   0.90 &  0.70 &    0.89 &    0.90 &    0.92 &    0.89 &    0.90 & $~$ 0.4 &    0.79 &    0.78 \\
   0.90 &  0.53 &    0.89 &    0.93 &    0.90 &    0.88 &    0.91 & -0.9 &    0.83 &    0.82 \\
   0.90 &  0.35 &    0.89 &    0.95 &    0.90 &    0.88 &    0.92 & -2.2 &    0.88 &    0.87 \\
   0.90 &  0.17 &    0.93 &    0.90 &    0.93 &    0.91 &    0.91 & -1.3 &    0.95 &    0.95 \\
\hline
   0.95 &  0.70 &    0.94 &    0.95 &    0.95 &    0.93 &    0.94 &  $~$ 0.9 &    0.83 &    0.82 \\
   0.95 &  0.53 &    0.95 &    0.94 &    0.96 &    0.94 &    0.95 & $~$ 0.5 &    0.88 &    0.87 \\
   0.95 &  0.35 &    0.97 &    0.97 &    0.97 &    0.95 &    0.97 & -2.5 &    0.93 &    0.92 \\
   0.95 &  0.17 &    0.96 &    0.94 &    0.95 &    0.94 &    0.95 & -0.2 &    1.01 &    1.01 \\
\hline
   1.00 &  0.70 &    1.03 &    1.01 &    1.02 &    1.02 &    1.02 & -1.9 &    0.88 &    0.88 \\
   1.00 &  0.53 &    1.02 &    1.02 &    1.01 &    1.01 &    1.02 & -2.2 &    0.93 &    0.93 \\
   1.00 &  0.35 &    0.99 &    0.99 &    0.99 &    0.98 &    0.99 & $~$ 0.9 &    0.99 &    0.98 \\
   1.00 &  0.17 &    1.02 &    1.02 &    1.00 &    1.01 &    1.02 & -1.8 &    1.07 &    1.07 \\
\hline
   1.05 &  0.70 &    1.03 &    1.06 &    1.03 &    1.03 &    1.04 & $~$ 0.5 &    0.93 &    0.92 \\
   1.05 &  0.53 &    1.04 &    1.06 &    1.03 &    1.04 &    1.05 & $~$ 0.0 &    0.99 &    0.98 \\
   1.05 &  0.35 &    1.06 &    1.06 &    1.04 &    1.05 &    1.06 & -0.8 &    1.05 &    1.04 \\
   1.05 &  0.17 &    1.06 &    1.05 &    1.04 &    1.05 &    1.05 & -0.4 &    1.14 &    1.13 \\
\hline
   1.10 &  0.70 &    1.09 &    1.10 &    1.07 &    1.09 &    1.09 & $~$ 0.6 &    0.99 &    0.98 \\
   1.10 &  0.53 &    1.09 &    1.06 &    1.07 &    1.09 &    1.07 & $~$ 2.4 &    1.05 &    1.04 \\
   1.10 &  0.35 &    1.10 &    1.09 &    1.07 &    1.10 &    1.10 & $~$ 0.4 &    1.12 &    1.11 \\
   1.10 &  0.17 &    1.12 &    1.12 &    1.09 &    1.12 &    1.12 & -2.1 &    1.21 &    1.20 \\
\hline
   1.15 &  0.70 &    1.12 &    1.12 &    1.10 &    1.13 &    1.12 & $~$ 2.4 &    1.06 &    1.04 \\
   1.15 &  0.53 &    1.13 &    1.14 &    1.11 &    1.14 &    1.14 & $~$ 1.3 &    1.12 &    1.11 \\
   1.15 &  0.35 &    1.17 &    1.16 &    1.13 &    1.18 &    1.17 & -1.4 &    1.19 &    1.18 \\
   1.15 &  0.17 &    1.14 &    1.18 &    1.11 &    1.15 &    1.16 & -1.0 &    1.27 &    1.26 \\
\hline
   1.20 &  0.70 &    1.17 &    1.17 &    1.15 &    1.20 &    1.17 & $~$ 2.6 &    1.13 &    1.12 \\
   1.20 &  0.53 &    1.21 &    1.21 &    1.17 &    1.23 &    1.21 & -0.9 &    1.19 &    1.19 \\
   1.20 &  0.35 &    1.18 &    1.23 &    1.16 &    1.22 &    1.21 & -0.5 &    1.27 &    1.26 \\
   1.20 &  0.17 &    1.20 &    1.19 &    1.16 &    1.22 &    1.20 & $~$ 0.3 &    1.34 &    1.32 \\
\hline
   1.25 &  0.70 &    1.23 &    1.26 &    1.21 &    1.24 &    1.24 & $~$ 0.7 &    1.19 &    1.19 \\
   1.25 &  0.53 &    1.28 &    1.26 &    1.23 &    1.25 &    1.27 & -1.5 &    1.27 &    1.27 \\
   1.25 &  0.35 &    1.26 &    1.24 &    1.23 &    1.25 &    1.25 & $~$ 0.0 &    1.34 &    1.36 \\
   1.25 &  0.17 &    1.28 &    1.27 &    1.23 &    1.25 &    1.27 & -2.0 &    1.47 &    1.46 \\
\hline
   1.30 &  0.70 &    1.32 &    1.29 &    1.28 &    1.28 &    1.30 & -0.2 &    1.26 &    1.27 \\
   1.30 &  0.53 &    1.31 &    1.33 &    1.28 &    1.28 &    1.32 & -1.7 &    1.34 &    1.36 \\
   1.30 &  0.35 &    1.33 &    1.31 &    1.29 &    1.28 &    1.32 & -1.6 &    1.44 &    1.47 \\
   1.30 &  0.17 &    1.31 &    1.32 &    1.27 &    1.27 &    1.31 & -0.9 &    1.59 &    1.60 \\

\hline
\end{tabular}
\end{table}

\subsection{Scaling relations in terms of $\numin_1$, $\braket{\Dnu}$ 
and $T_{\rm eff}$}
{ In previous studies, the main asteroseismic parameters 
used to infer the fundamental stellar properties have been
$\braket{\Dnu}$  and $\numax$. }
If we have high quality data, then one can also extract the average 
small {frequency} separation.
In addition to these, $\nu_{\rm min1}$ and \numin$_2$ increase 
asteroseismic ability to predict stellar properties.
  
$\nu_{\rm min1}$ and \numin$_2$ can easily be determined from oscillation 
frequencies. 
We show above that $\nu_{\rm min1}/M$ is equivalent to $\nu_{\rm max}$ in 
new scaling relations. 
Then, equation (1) can be written in terms of $\nu_{\rm min1}$ as 
\begin{equation}
\frac{M}{{\rm M}_{\sun}}=\left( \frac{(\numin_1/\nu_{\rm min1\sun})^3}{(\braket{\Delta \nu}/\braket{\Delta \nu_{\sun}})^4}\left( \frac{T_{\rm eff}}{\rm T_{eff\sun}}\right)^{1.5}\right)^{1/4}
\label{equ:Mmin}
\end{equation}
The masses computed from equation (11) { are plotted against 
model masses in Fig. 10.}
The agreement is good between the { two mass estimates.
There is a very slight deviation from a linear relationship.
For better agreement, the power in the right-hand side of 
equation (11) should be modified to}
\begin{figure}
\includegraphics[width=97mm,angle=0]{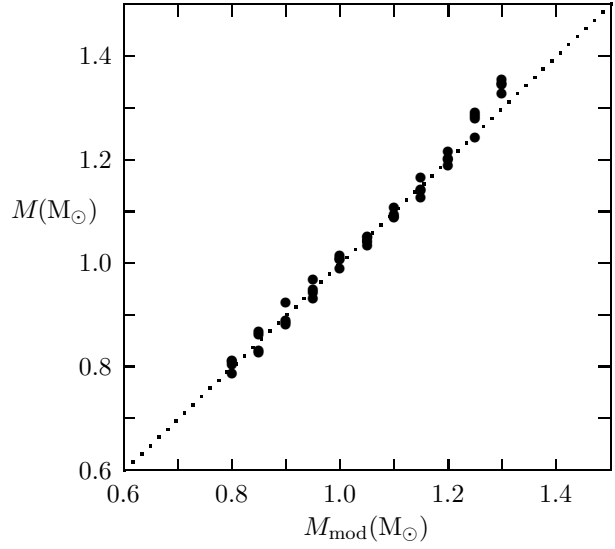}
\caption{The computed mass (equation 11) from oscillation frequencies 
with respect to model mass.
There is a slight difference for 1.25 and 1.30 \MS models.
}
\end{figure}
\begin{equation}
\frac{M}{{\rm M}_{\sun}}=\left( \frac{(\numin_1/\nu_{\rm min1\sun})^3}{(\braket{\Dnu}/\braket{\Delta \nu_{\sun}})^4}\left( \frac{T_{\rm eff}}{\rm T_{eff\sun}}\right)^{1.5}\right)^{1/4.26}.
\label{equ:Mmin}
\end{equation}
The computed masses  ($M_{\rm e12}$) from equation (12) are also listed in Table 1.
The maximum difference between equation (12) and model mass is about 
2.5 per cent (see Fig. 14 in Section 5.9). 
{ In the mass interval for solar-like oscillating stars 
near the MS, two structural transitions occur.
While the CZ becomes shallow in the outer regions as stellar mass 
increases, a convective core develops in the central region. }
Therefore two separate fits may in turn be required (see below).  


If we insert the expression we derived for $\numax$ in equation (2),
\begin{equation}
\frac{R}{{\rm R}_{\sun}}=\frac{{\rm M}_{\sun}}{M}\frac{\numin_1/\nu_{\rm min1\sun}}{(\braket{\Dnu}/\braket{\Delta \nu_{\sun}})^2}\left( \frac{T_{\rm eff}}{\rm T_{eff\sun}}\right)^{0.5}
\label{equ:Rmin}
\end{equation}
is obtained for radius. { We insert equation (11) in equation (13) and then find}
\begin{equation}
\frac{R}{{\rm R}_{\sun}}=\frac{(\numin_1/\nu_{\rm min1\sun})^{1/4}}{(\braket{\Dnu}/\braket{\Delta \nu_{\sun}})}\left( \frac{T_{\rm eff}}{\rm T_{eff\sun}}\right)^{1/8}.
\label{equ:Rminp}
\end{equation}
{ The uncertainty in the above expression is 3.5 per cent.
In order to {raise} this uncertainty 
we plot a figure similar to Fig. 10 but for radius. 
{ 
We obtain more precise results than given {by equation (14) if we reduce 
the right-hand side of equation (14) to the power of 0.95.}
}
 
 
}
\subsection{Mass and radius in terms of $\numin_1$ and $\braket{\Delta \nu}$}
{ The $T_{\rm eff}$ values of many $Kepler$  target stars 
are not determined very precisely. }
If we assume a typical uncertainty $\Delta T_{\rm eff} \approx 200$ K,
the uncertainty is about 3 per cent for $T_{\rm eff}=6000$ K. 
This causes an uncertainty in $M$ about 4 to 5 per cent.
To reduce this uncertainty in $M$, { here we try} to derive 
expressions for $M$ and other fundamental properties in terms of 
{ purely asteroseismic quantities 
$\braket{\Delta \nu}$ and \numin$_1$.
These simple relations are obtained to illustrate 
the diagnostic potentials of new asteroseismic parameters (\numin$_1$).
They are not the final forms that one can derive.}

For a lower uncertainty, two separate formula may be derived for two mass 
intervals $1$$-1.2$ and  $1.2$$-1.3$ \MSbit.
If  $M < 1.2 $ M$_{\sun}$, then 
\begin{equation}
\frac{M}{M_{\sun}}=\frac{(\numin_1/\nu_{\rm min1\sun})^{0.92}}{(\braket{\Dnu}/\braket{\Delta \nu_{\sun}})^{1.27}}.
\label{equ:Mmin}
\end{equation}
If $M > 1.2 $ M$_{\sun}$, then
\begin{equation}
\frac{M}{M_{\sun}}=1.134 \frac{(\numin_1/\nu_{\rm min1\sun})^{0.35}}{(\braket{\Dnu}/\braket{\Delta \nu_{\sun}})^{0.47}}.
\label{equ:Mmin}
\end{equation}
{ The uncertainties in equations (15) and (16) are less than 
2 per cent; see Table 1.}


Again using only the asteroseismic quantities $\braket{\Delta \nu}$ 
and \numin$_1$ ~ we try to obtain {an} expression for 
stellar radius. 
Indeed, many relations can be found by { similar fitting 
procedures;} the most precise one we obtain is  
\begin{equation}
\frac{R}{{\rm R}_{\sun}}=\left(\frac{\numin_1}{\nu_{\rm min1\sun}}\right)^{0.23}\left(\frac{\braket{\Delta \nu_{\sun}}}{\braket{\Dnu}}\right)^{0.99}.
\label{equ:Rmin}
\end{equation}
The maximum difference between equation (17) and model radius is  1.5 per cent.  
{ Similarly, we also derive an expression for gravitational 
acceleration at the stellar surface ($g$): }
\begin{equation}
\frac{g}{{\rm g}_{\sun}}=\left(\frac{\numin_1}{\nu_{\rm min1\sun}}\right)^{0.48}\left(\frac{\braket{\Dnu}}{\braket{\Delta \nu_{\sun}}}\right)^{0.78}.
\label{equ:Rmin}
\end{equation}
{ Equation (18) is also a very { precise} relation.} 
It is in very good agreement with the model $g$ values. 
The maximum difference between them is less than 2 per cent.

\subsection{Effective temperature and luminosity  in terms of $\numin_1$, $\numax$  and $\braket{\Delta \nu}$}
$T_{\rm eff}$ is one of the very important stellar parameters and 
{ it is} not precisely determined in many cases.
Luminosity, however, is one of the essential parameters if one compares 
stellar models with stars.
{ It is the most rapidly changing parameter throughout 
MS evolution and therefore is considered as an age indicator.}
In order to show the diagnostic potential of asteroseismic properties, 
we also derive fitting formula for $T_{\rm eff}$ and luminosity $L$.
{ For $T_{\rm eff}$ of models with $M>1.0 \MSbit$, }
\begin{equation}
\frac{T_{\rm eff}}{\rm T_{eff\sun}}=\frac{(\numin_1/\nu_{\rm min1\sun})^{0.26}}{(\braket{\Dnu}/\braket{\Delta \nu_{\sun}})^{0.4}}.
\label{equ:Rmin}
\end{equation}
The maximum difference between equation (19) and $T_{\rm eff}$ of the 
models is $150$ K, but the mean difference is about $50$ K.
The method for determination of $T_{\rm eff}$ from $\Delta n_{\rm x1}$ gives 
much more precise results {(see Fig. 6). 
}

The fitting formula obtained for luminosity as a function of the 
asteroseismic parameters is
\begin{equation}
\frac{L}{{\rm L}_{\sun}}=2.06\frac{\numin_1}{\nu_{\rm min1\sun}}\frac{\braket{\Delta \nu_{\sun}}}{\braket{\Dnu}}\frac{\nu_{\rm max\sun}}{\numax}-1.01.
\end{equation}
In Fig. 11, the luminosity derived from the asteroseismic parameters 
is plotted with respect to the model luminosity. 
The agreement seems excellent at least for the MS models with solar composition.
This result is very impressive because luminosity is one of the most 
uncertain stellar parameter derived from observations.

\begin{figure}
\includegraphics[width=97mm,angle=0]{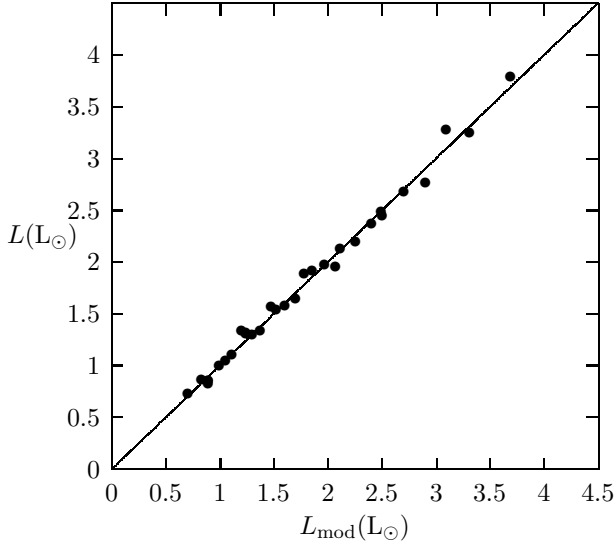}
\caption{Luminosity derived from oscillation frequencies ({ equation 20}) with 
respect to model luminosity.
}
\end{figure}

\subsection{Age in terms of $\numin_1$, $\braket{\Delta \nu}$ and $\braket{\delta \nu_{02}}$}
\begin{figure}
\includegraphics[width=97mm,angle=0]{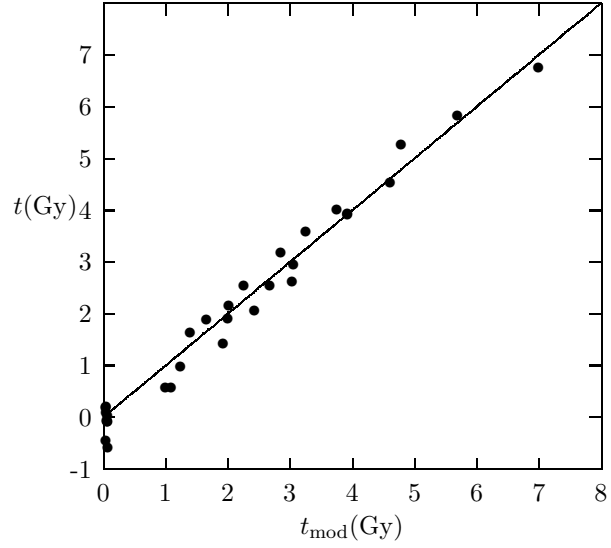}
\caption{Age derived from oscillation frequencies ({ equation 22}) with 
respect to model age.
}
\end{figure}
{ The age} of a star is one of the most difficult parameters 
to compute. 
It is very sensitive function of stellar properties,
such as mass and chemical composition. 
The number of stars {for which} we know these properties is unfortunately very small.
Therefore the { promise of asteroseismology to better
constrain} stellar age is very important. 
The mean value of small { frequency} separation, 
$\braket{\delta \nu_{02}}$, is a very good age indicator.  
We obtain a fitting formula for the stellar age, which is a function of  
$\numin_1$, $\numax$,  $\braket{\Dnu}$ and $\braket{\delta \nu_{02}}$.

{ Defining}
\begin{equation}
r_\nu=\left(\frac{\numax}{\nu_{\rm max\sun}} 
\frac{\braket{\Dnu}}{\braket{\Delta \nu_{\sun}}}\right)^{0.7},
\end{equation}
we obtain the fitting formula for the stellar age as
\begin{equation}
t({\rm Gyr})=6.93 \left(r_\nu(0.91-\frac{\braket{\delta \nu_{02}}}{\braket{\delta \nu_{02\sun}}})
+ 2.96 - 2.18\frac{M}{{\rm M}_{\sun}} \right)^{1.18}, 
\end{equation}
where $M/{\rm M}_{\sun}$ is computed from equation (10) and therefore 
a function of $\numin_1$ and $\numax$. 
The ages computed from equation (22) are plotted with respect to model age 
in Fig. 12. 
For some ZAMS models, equation (22) based on the asteroseismic parameters 
gives negative values for the age. 
Age in such a case is considered to be very small and can be taken as
the ZAMS age. 
For the other models, the difference between the age derived from 
equation (22) and model age is less than 0.5 Gyr. 
For the Sun, { equation (22)} gives its age as 4.8 Gyr.  
{ This result is in very good agreement with the solar age found by 
Bahcall, Pinsonneault \& Wasserburg (1995), 4.57 Gyr.} 
One should notice that the { models are constructed} 
with solar composition. 
The effect of metallicity on these relations (and chemical composition in 
general, for example effect of $X$) { should be studied further.}

%

\subsection{Asteroseismic diagram for stellar structure and evolution in 
terms of $\braket{\Dnu}$ and $\numin_1$}
\begin{figure}
\includegraphics[width=97mm,angle=0]{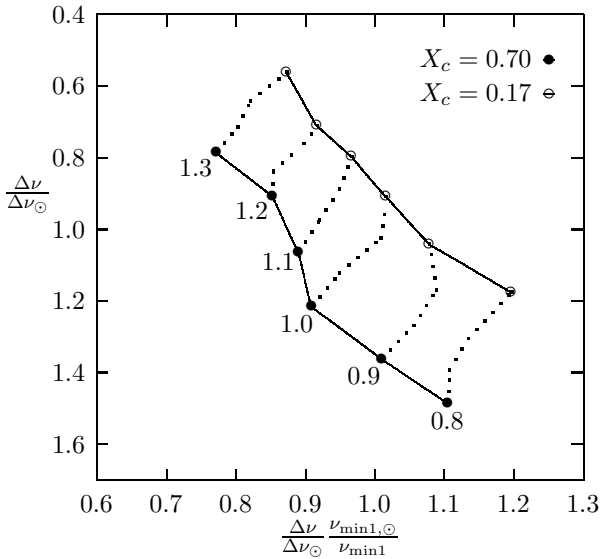}
\caption{ AD for solar-like oscillating stars. 
The numbers show the model masses in units of solar mass. 
This form of the diagram is compatible to the classical HR diagram.
}
\end{figure}

{ After the pioneering study of Christensen-Dalsgaard (1988) 
on the seismic  Hertzsprung-Russel (HR) diagram}, many papers have 
appeared in the literature on this subject (e.g., Mazumdar 2005; 
Tang, Bi \& Gai 2008; White et al. 2011). 
In Christensen-Dalsgaard (1988), the vertical and horizontal axes are the 
small and large separations
between the oscillation frequencies, respectively. 
{ The uncertainty in the large separation depends on the mean
uncertainty in frequencies.} 
For the small separations, however, the situation is a bit different. 
$\braket{\delta \nu_{02}}$ depends also on which interval of $n$ is 
{ used, and it is not very certain in many cases.}

We suggest a new asteroseismic diagram (AD) in terms of 
$\braket{\Dnu}$ and $\numin_1$ in Fig. 13. 
{ The vertical axis is $\braket{\Dnu}$ in { the new AD and 
the} horizontal axis is 
chosen as $\braket{\Dnu}/\numin_1$ in solar ıunits.  
{ This form of the AD is very compatible with the classical 
HR diagram.}
The ZAMS line is in left-hand { part and TAMS line is in the 
right-hand part of the AD. }
Furthermore, the low mass models { appear in the lower part
and high-mass models are in the upper part of the AD.}
Thus, evolutionary tracks of stars in the HR diagram and the AD are very 
compatible with each other.
}

\subsection{The effect of metallicity on the relation between stellar mass 
and oscillation frequencies}
{
Eigenfrequencies of a model depend on many stellar parameters. 
This can lead us to expect that metallicity may influence the relations 
we derive in the present study. 
Equation (10), for example, can be rewritten as

\begin{equation}
\frac{M}{\MS}=\frac{\nu_{\rm min1}}{\nu_{\rm min1\sun}}\frac{\nu_{\rm max\sun}}{\nu_{\rm max}} 
\left(\frac{Z}{{\rm Z}_{\sun}}\right)^{\beta_z},
\end{equation}
where $\beta_z$ is the parameter to be determined from model computations. 
Our preliminary results show that $\beta_z$ is about 0.1. 
}
\subsection{On the uncertainty in \numin$_1$  and relations between 
asteroseismic quantities and fundamental stellar parameters}
\begin{figure}
\includegraphics[width=97mm,angle=0]{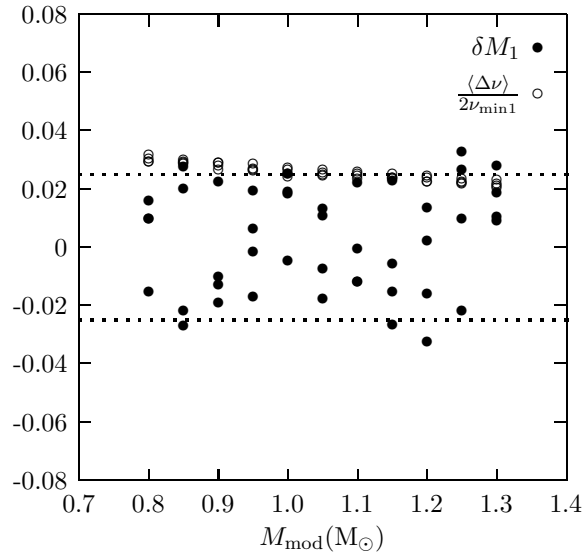}
\caption{  
 The mass difference between $M_1=1.188 \numin_1/\numax$ and model mass (filled circles)
is plotted with respect to $M_{\rm mod}$. The uncertainty in $M_1$ is about 0.025 \MSbit.  
$\braket{\Delta \nu}/(2 \numin_1)$ is also about 0.025. This implies that uncertainty in $M_1$ 
depends on how accurate $\numin_1$ is. The uncertainty in $\numin_1$ is about $\braket{\Delta \nu}/2$.
}
\end{figure}
{ 
The main uncertainty in our results comes from uncertainty in \numin$_1$. 
As an example,
we plot the difference between the model mass and mass derived from 
$M_1=1.188 \numin_1/\numax$ ($\delta M=M_1-M_{\rm mod}$) with respect to 
model mass (in solar units) in Fig. 14. 
The mean difference is negligibly small and about 0.0026 \MSbit.
However, the differences for the range 0.8-1.3 \MS is mostly less than 
0.025 \MSbit. 
This must be due to determination of \numin$_1$. 
For comparison, $\braket{\Delta \nu}/(2 \numin_1)$ is also plotted. 
It is also about 0.025. 
This implies that \numin$_1$ is { uncertain by} about $\braket{\Delta \nu}/2$. 
This amount of uncertainty seems reasonable { considering} our
method for determination of \numin$_1$.

%

}
\section{Conclusion}
In the present study, 
we analyse two frequencies ($\numin_1$ and $\numin_2$) 
at which ${\Dnu}$ is { minimized.}
{ These frequencies correspond to the modes whose one of radial 
displacement peaks coincide with the minimum of $\Gamma_1$ in 
the He {\small II} ionization zone (see Fig. 7).
They have very strong diagnostic potential}. 
If we divide any of them by the frequency of maximum amplitude ($\numax$) 
we find stellar mass very precisely. 
In the previous expressions in {the} literature, $M$ is 
found in terms of $\numax$, ${\Dnu}$ and $T_{\rm eff}$. 
The { precision of stellar mass found from asteroseismic methods depends 
 the precisions of the inferred frequencies 
($\numin_1$, $\numin_2$ and $\numax$). }


Both $\numin_1$ and $\numin_2$ are functions of stellar mass and age 
{ in particular, and in general depend on}
all the parameters influencing stellar structure. 
Such dependences in some respects complicate the situation, 
{but} they become very strong tools if the relations 
between parameters and frequencies are well-constructed.
{ Variations of both $\numin_1$ and $\numax$ with evolution} are the same and therefore 
their ratio remains constant and {yields} stellar mass.

The method we find is in principle very precise. 
Fundamental properties, such as mass, radius, gravity and $T_{\rm eff}$ 
are determined within the { precision} of 2 to 3 per cent. 
This is the level of accuracy for the well-known eclipsing binaries. 
We also derive a fitting formula for luminosity (equation 20) and 
age (equation 22) as functions of asteroseismic quantities. 

{ Frequencies} $\numin_1$ and $\numin_2$ are equivalent 
to each other. 
{ They obey the same relations, at least for the MS stars. 
{ The { mode order} difference  between them ($(\numin_1-\numin_2)/\Dnu$) is 
related to the depth of the CZ.} 
However, the mass of {the} CZ is best given by 
any of the { mode order} differences 
$\Delta n_{\rm x1}=(\numax-\numin_1)/\Dnu$ or 
$\Delta n_{\rm x2}=(\numax-\numin_2)/\Dnu$.
For example, $M_{\rm CZ}=0.066\Delta n_{\rm x1}\MS$ (see Fig. 6).  
$\Delta n_{\rm x1}$ and $\Delta n_{\rm x2}$ are also 
very important tools for precise determination of $T_{\rm eff}$.}

We obtain scaling relations using asteroseismic quantities, 
$\numin_1$, \Dnu, $\numax $ and $T_{\rm eff}$.
{ We also derive expressions for fundamental stellar parameters 
by { eliminating} $T_{\rm eff}$.    }

We also suggest a new AD. 
The $y$-axis is the large separation  $\Dnu$ and $\Dnu/\numin_1$ is the $x$-axis.
In this form of { the $x$-axis, the} evolutionary tracks 
and ZAMS and TAMS lines in AD are compatible to those in 
the traditional HR diagram. 

{ The present study is essentially based on the oscillation frequencies 
of models with solar composition.
Our preliminary results on the models with higher metallicities than the 
solar metallicity show that relations between asteroseismic quantities 
and fundamental stellar parameters are changing with the metallicity. 
A similar test should be carried out for { variations in 
the hydrogen abundance.}
}
In the next paper of this series of papers, we will do this test and apply 
the methods developed in the present study to the 
$Kepler$  and $CoRoT$  data and also test the effects of chemical composition.
 
\section*{Acknowledgements}
{ This paper is dedicated to Professor D. Ezer-Eryurt. 
The original version of the {\small ANK\.I} stellar evolution code is developed
by Dr Ezer-Eryurt and her colleagues. 
Dr Ezer-Eryurt 
{ passed away on 2012 September 13 after 
leaving excellent memories behind her.}
The authors of this paper are grateful to her.
Enki is the god of knowledge in Summerians mythology and Anki means universe.
}
{ The anonymous referee and Professor Chris Sneden  are acknowledged 
for their suggestions which improved the presentation of the manuscript.}
We would like to thank Sarah Leach-Laflamme for her help in checking the 
language of the early version of the manuscript.
This work is supported by the Scientific and Technological Research 
Council of Turkey (T\"UB\.ITAK: 112T989).
{ 
During the second revision of this manuscript, the updated code {\small ANK\.I} is completely deleted
by the system administrator of fencluster without any warning. 
The academic life here in our 'lonely and beautiful country' is 
very difficult, but not hopeless.   
}

\newpage
\onecolumn
\appendix
\section[t]{Online-only figures for Comparison of asteroseismic inferences with the model values
           }
\twocolumn

\begin{figure}
\includegraphics[width=97mm,angle=0]{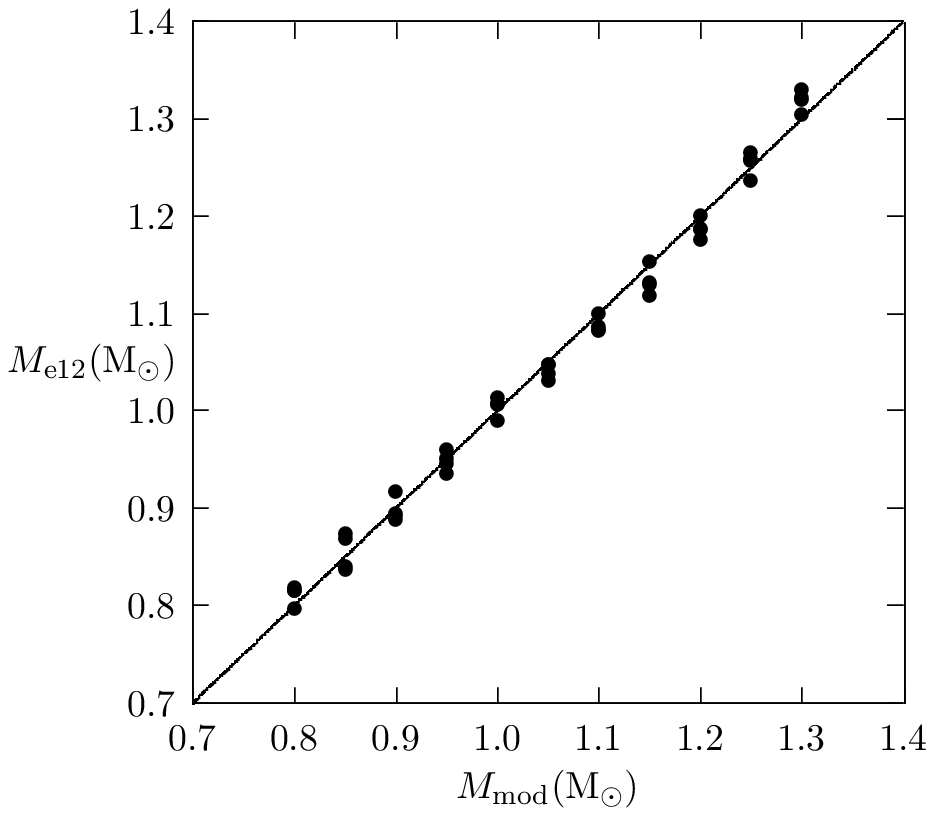}
\caption{ The asteroseismic mass computed from equation (12) ($M_{\rm e12}$) is plotted with respect to 
model mass ($M_{\rm mod}$).
}
\end{figure}
\begin{figure}
\includegraphics[width=97mm,angle=0]{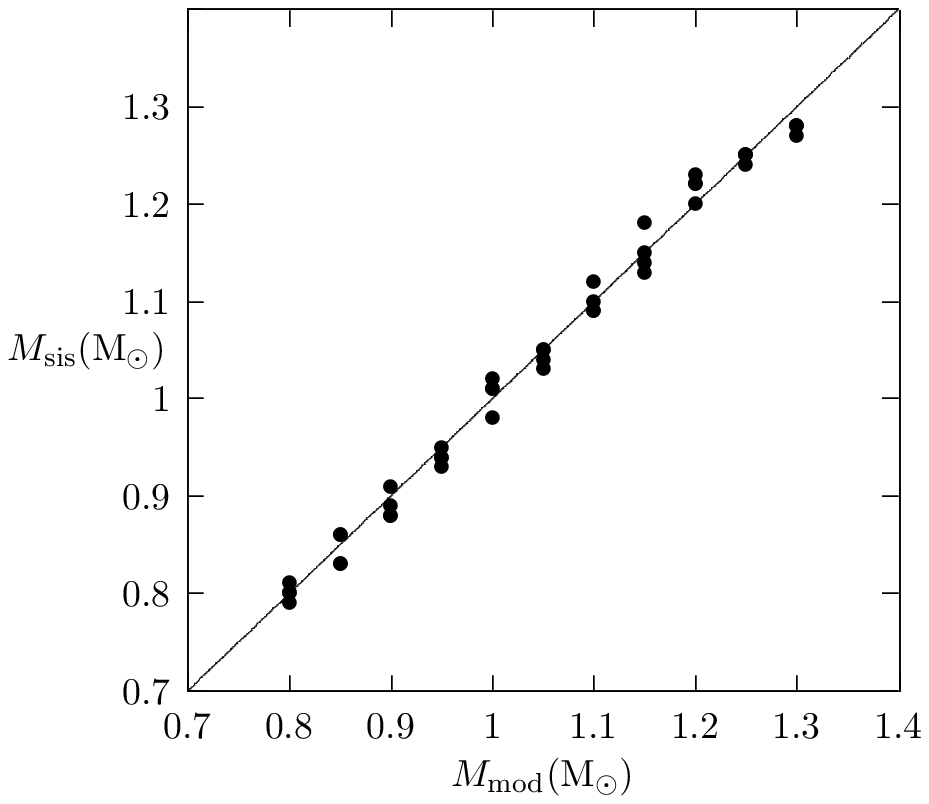}
\caption{ 
The asteroseismic mass $M_{\rm sis}$ computed from equation (15) or (16) is plotted with respect to 
$M_{\rm mod}$.
}
\end{figure}
\begin{figure}
\includegraphics[width=97mm,angle=0]{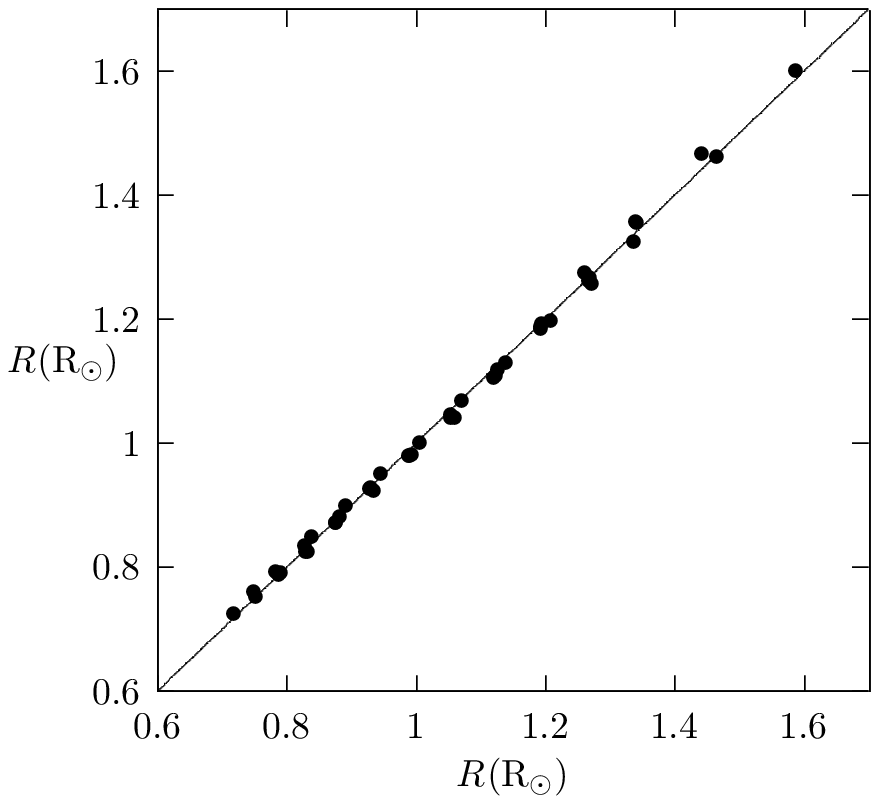}
\caption{ 
The asteroseismic radius $R_{\rm e17}$ computed from equation (17) is plotted with respect to 
model radius ($R_{\rm mod}$).
}
\end{figure}
\begin{figure}
\includegraphics[width=97mm,angle=0]{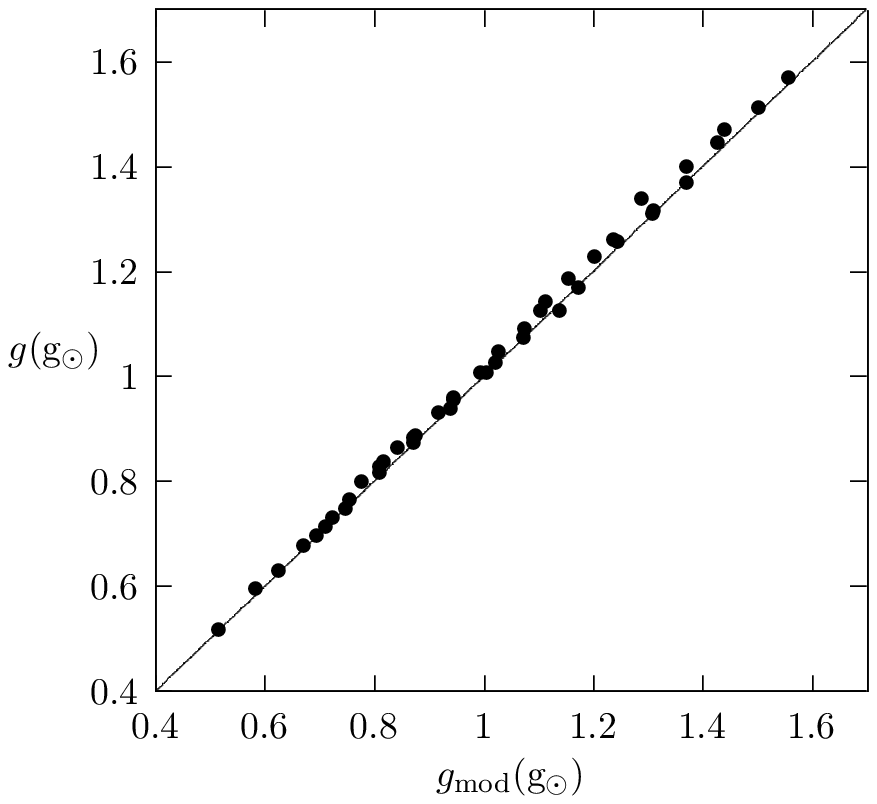}
\caption{ 
The asteroseismic surface gravitational accelaration $g_{\rm}$ computed from equation (18) is plotted 
with respect to model accelaration ($g_{\rm mod}$).
}
\end{figure}

\begin{figure}
\includegraphics[width=97mm,angle=0]{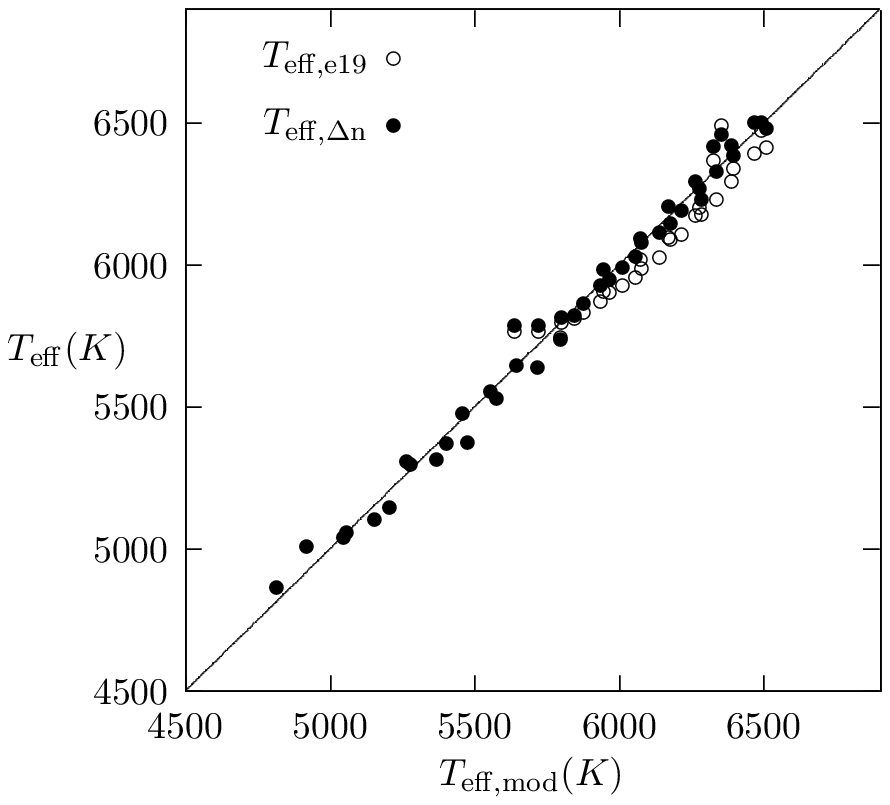}
\caption{ 
The asteroseismic effective temperatures computed from equation (19) (circle) and 
$T_{{\rm eff},\Delta n}=(1.142-9.63~10^{-3}(\Delta n_{\rm x1}+4)^{1.35}) {\rm T_{eff \sun}}$ 
(filled circle) are plotted 
with respect to model $T_{\rm eff}$.
}
\end{figure}

\label{lastpage}
\end{document}